\documentclass[runningheads]{llncs}

\usepackage{mypackages}
\usepackage{mycommands}

\title{Post-Quantum Oblivious Transfer from Smooth Projective Hash Functions with Grey Zone}
\titlerunning{Post-Quantum \OT from SPHF with Grey Zone}

\author{Slim Bettaieb\inst{1} \and Loïc Bidoux\inst{2} \and Olivier Blazy\inst{3} \and Baptiste Cottier\inst{4} \and David Pointcheval\inst{4}}
\institute{Worldline, France
\and Technology Innovation Institute, United Arab Emirates
\and Ecole Polytechnique, IPP, France
\and DIENS, CNRS, ENS/PSL, Inria, Paris, France}
\date{}
\authorrunning{Bettaieb, Bidoux, Blazy, Cottier, Pointcheval}
\begin{document}
\maketitle

\begin{abstract}
    Oblivious Transfer (\OT) is a major primitive for secure multi-party computation.
    Indeed, combined with symmetric primitives along with garbled circuits, it allows any secure function evaluation between two parties.
    In this paper, we propose a new approach to build \OT protocols.
    Interestingly, our new paradigm features a security analysis in the Universal Composability (\UC) framework and may be instantiated from post-quantum primitives.
    In order to do so, we define a new primitive named Smooth Projective Hash Function with Grey Zone (\SPHFwG) which can be seen as a relaxation of the classical Smooth Projective Hash Functions, with a subset of the words for which one cannot claim correctness nor smoothness: the grey zone.
    As a concrete application, we provide two instantiations of \SPHFwG respectively based on the Diffie-Hellman and the Learning With Errors (LWE) problems.
    Hence, we propose a quantum-resistant \OT protocol with \UC-security in the random oracle model.
\end{abstract}

\section{Introduction}
Smooth Projective Hash Function (\SPHF), or Hash Proof System as introduced by Cramer and Shoup in \cite{EC:CraSho02}, is a cryptographic primitive initially designed to provide IND-CCA encryption schemes.
Over the years, {\SPHF}s have been used for many applications such as Password-Authenticated Key Exchange \cite{EC:GenLin03,C:AbdChePoi09,TCC:KatVAi11,PKC:BBCPV13}, Zero-Knowledge Proofs \cite{PKC:JutRoy12,C:BBCPV13} or Witness Encryption \cite{EPRINT:DerSla15b}.
Since their introduction, {\SPHF}s have been developed over classical hard problems such as discrete logarithm or factorization. However, post-quantum cryptography does not seem to be as easily compliant with \SPHF. In \cite{AC:KatVai09a}, Katz \etal introduced \emph{Approximate Smooth Projective Hash Functions}. The correctness property of an \SPHF claims that the hash value and the projective hash value are required to be equal on words in an NP-language, when knowing a witness, while the smoothness property expects them to be independent when no witness exists. Approximate \SPHF uses an approximate correctness, that allows those values to be close, relatively to a given distance. Furthermore, languages relying on code-based or lattice-based ciphertexts result in a gap between the set of valid ciphertexts of a given value $\mu$, and the values that decrypt into $\mu$. As mentioned in \cite{EPRINT:BBBCG21}, an adversary could maliciously generate one of those ciphertexts and open the door for practical attacks. The presence of this gap can also be problematic when expecting to work in the Universal Composability framework \cite{FOCS:Canetti01}.

\paragraph{Related Works.} In code-based cryptography, the first proposition was made by Persichetti in \cite{PQCRYPTO:Persichetti13}. The \SPHF proposed there uses a weaker smoothness definition, called universality. Strictly speaking, this is not a drawback as we can transform an \SPHF with universality property to a word-dependent \SPHF with smoothness property. However, the main issue with this candidate is that the proof is done on random keys, rather than the whole keys. This has for consequence that an adversary can exploit some well-chosen keys resulting in a failure of the proof. A second construction was designed in \cite{EPRINT:BBBCG21}. As said before, when working with lattices and codes, languages based on ciphertexts present a grey zone. In this work, Bettaieb \etal withdraw this gap using a zero-knowledge proof asserting if two different ciphertexts of the same message are valid, reducing the \SPHF on the set of valid ciphertexts, resulting in the first \emph{gapless} post-quantum \SPHF. A solution based on codes is also given in \cite{ShoAre21}, but their solution offers an Approximate \SPHF with computational smoothness, while real \SPHF expects statistical/perfect smoothness.  In lattice-based cryptography, the first construction was given in \cite{AC:KatVai09a} where Katz \etal introduced the notion of Approximate \SPHF. Their language not being exactly defined as the valid LWE-ciphertexts, decoding procedure was expensive, as detailed in \cite{PKC:BBDQ18}. This latter article, motivated by this issue, offers the first non-approximated \SPHF based on lattices. While the two previous constructions are in the standard model, Zha \etal \cite{AC:ZhaYu17} propose a \SPHF requiring access to a random oracle. Indeed, their language relies on simulation-sound non-interactive zero-knowledge proofs, that we are not able to construct efficiently without random oracles.

\paragraph{Contribution.} As mentioned above, a gap appears when working with cryptography based on lattices or codes. Rather than withdraw this gap as done in \cite{EPRINT:BBBCG21}, we focus on the requirements needed in order to tame this gap, with an additional notion of \emph{Decomposition Intractability} when trying to exploit this gap. 
Therefore, we introduce \emph{Smooth Hash Projective Functions with Grey Zone} (\SPHFwG) as an \SPHF with the \emph{Decomposition Intractability} property: we will require a language $\cL$, hard to decide, as for any non-trivial \SPHF, but also with additional intractability for finding two \emph{complementary} words in $\cL$ or the gap.
As an application of \SPHFwG, we show that one can design an Oblivious Transfer from any \SPHF with Grey Zone on languages of ciphertexts for homomorphic encryption, where the security relies on the semantic security. 

We provide two concrete instantiations of \SPHFwG: the first one relies on the Diffie-Hellman Problem and the ElGamal cryptosystem. As no decryption failure occurs with the ElGamal cryptosystem, the grey zone is empty and the decomposition intractability is obvious. One can note that the resulting \SPHFwG is \emph{de facto} an \SPHF. The idea behind this instantiation is, on the one hand, to familiarise the reader with our construction, and on the other hand, to point out the fact that the construction is also available from any classical \SPHF. A second instantiation is based on lattices and more precisely from the \emph{Learning with Errors} problem. This allows to underline the genericity of our framework.


\section{Preliminaries}
\label{sec:preliminaries}
\subsection{Oblivious Transfer}
Oblivious transfer, introduced by Rabin~\cite{Rabin81},
involves a sender with input two messages $m_0, m_1$ and a receiver with input a selection bit $b$. so that the latter receives $m_b$ and nothing else, while the former does not learn anything. It provides sender-privacy (no information leakage about $m_{1-b}$) and receiver-privacy (no information leakage about $b$). 


\subsection{Universal Composability}
Universal Composability is a security model taking into account the whole environment (i.e. all exterior interactions) of the execution. Concretely, if a protocol is proven to be universally composable (or \emph{UC-secure}), it can be used concurrently with other protocols without compromising the global protocol security. Proving universally composable security is done thanks to the real world / ideal world paradigm. In the ideal world, we consider an access to a trusted third party. A protocol $\varPi$ is \emph{UC-secure}, if, for all environment $\cE$, there exists a simulator $\cS$ such that the execution of the protocol $\varPi$ with adversary $\cA$ in the real world, is indistinguishable with the execution of the functionality $\cF$ with simulator $\cS$ in the ideal world.

\subsection{Smooth Projective Hash Functions}
Introduced in 2002 \cite{EC:CraSho02}, Smooth Projective Hash Functions (\SPHF), also known as Hash Proof System (\textsf{HPS}), initially aim to build the first public key encryption scheme secure against chosen ciphertext attacks. 
Nowadays, \SPHF are mainly used for Honest Verifier Zero Knowledge Proofs or Witness Encryption.
Such functions work on NP-languages $\cL\subset\cX$, defined by a binary relation $\cR$ such that for any word $x\in\cX$, $x \in \cL$ if and only if there exists a witness $w$ such that $\cR(x,w)=1$. Then, an \SPHF defined on $\cL \subset \cX$ with values in $\cV$ is defined by five algorithms:
\begin{itemize}
    \item $\setup(1^\kappa)$: Generates the parameters $\param$ from $\kappa$, the security parameter. $\param$ includes a description of $\cL$, a language in $\cX$;
    \item $\HashKG(\param)$: Generates a random hash key $\hk$;
    \item $\ProjKG(\hk)$: Derives the projection key $\hp$;
    \item $\Hash(\hk, x)$: Returns the hash value $H_\hk \in \cV$ associated to the word $x$;
    \item $\ProjHash(\hp, x, w)$: Returns $H_\hp \in \cV$ using a witness $w$ linked to the word $x$. 
\end{itemize}
Those algorithms should ensure two requirements:
\begin{itemize}
  \item \textbf{Correctness:} For any $x \in \cL$, with witness $w$, $H_\hk=H_\hp$ where, with the keys $\hk \gets \HashKG(\param)$, $\hp \gets \ProjKG(\hk)$, the hash values are $H_\hk \gets \Hash(\hk, x)$ and $H_\hp \gets \ProjHash(\hp, x, w)$, under the condition that $\cR(x,w)=1$;
  \item \textbf{Smoothness:} For any $x \in \cX \backslash \cL$, the distributions of $(\hp,H_\hk)$ and $(\hp,v)$ are indistinguishable where, for the keys $\hk \gets \HashKG(\param)$ and $\hp \gets \ProjKG(\hk)$, $H_\hk \gets \Hash(\hk, x)$, and $v\getsr \cV$;
\end{itemize}
The aforementioned definition of smoothness was introduced by Cramer and Shoup in \cite{EC:CraSho02}. 
Two variants of this definition have later been proposed:
%
%
%
%
%
%
%
%
%
%
%
The first variation has been provided by Gennaro and Lindell in \cite{EC:GenLin03}, leading to the notion of \GL-\SPHF. 
The only difference with the definition of Cramer and Shoup (recalled above) is that the projection key $\hp$ may depend on the word $w$ of the language.  
The second variant, introduced by Katz and Vaikuntanathan in \cite{AC:KatVai09a} considers the ability for an attacker to maliciously generate the word $w$ after seeing the projection key $\hp$.
In \KV-\SPHF, the projection depends only on the hashing key and ensures the smoothness even if the word $w$ is chosen after having seen the projection key.
\GL-\SPHF will be enough for our applications, with word-dependent projection keys, as the word will be known beforehand.




\section{Smooth Projective Hash Functions with Grey Zone}
\label{sec:SPHFwGZ}
Our first contribution is the formalization of Smooth Projective Hash Functions with a Grey Zone ($\SPHFwG$) which is a relaxation of the classical \SPHF in which one cannot claim correctness nor smoothness for a subset of the words. Later, we will provide a quantum-resistant $\SPHFwG$ based on lattices.
With this new definition, we will have two disjoint languages $\cL,\cL'\subset\cX$ that will not necessarily partition the superset $\cX$: the remaining subset $\cX \backslash (\cL \cup \cL')$ will be the grey zone.

\subsection{Basic Definitions}
Let us describe our relaxation of \emph{Smooth Projective Hash Function} from~\cite{C:CraSho98} to encompass a \emph{Grey Zone}. An $\SPHFwG$ is defined with a tuple of algorithms:
\begin{itemize}
  \item $\setup(1^\kappa)$: Generate the parameters $\param$ from $\kappa$, the security parameter, or an explicit random tape $(\sigma,\rho)$ in $\cS_0\times\cR_0$. $\param$ includes a description of $\cL, \cL', \cX$, where $\cL \cup \cL' \subset \cX$ and $\cL \cap \cL' = \emptyset$, and $\cL$ is a language hard to decide in $\cX$;
  \item $\HashKG(\param)$: Generates a random hash key $\hk$;
  \item $\ProjKG(\hk,x)$: Derives the projection key $\hp$ (it may need $x$ as input);
  \item $\Hash(\hk, x)$: Returns the hash value $H_\hk \in \cV$, where $\cV$ is the set of hash values, associated to the word $x$;
  \item $\ProjHash(\hp, x, w)$: Returns $H_\hp \in \cV$ using a witness $w$ linked to the word $x$. 
\end{itemize}
%
As the classical \SPHF, our \SPHFwG verifies the following statistical properties, for any setup execution that provides $\param$, defining $\cL,\cL',\cX$:
\begin{itemize}
  \item \textbf{Correctness:} For any $x \in \cL$, $H_\hk=H_\hp$, where $\hk \gets \HashKG(\param)$, $\hp \gets \ProjKG(\hk,x)$, $H_\hk \gets \Hash(\hk, x)$, and $H_\hp \gets \ProjHash(\hp, x, w)$ for the witness $w$ of $x \in \cL$;
  \item \textbf{Smoothness:} For any $x \in \cL'$, the distributions of $(\hp,H_\hk)$ and $(\hp,v)$ are indistinguishable, where $\hk \gets \HashKG(\param)$, $\hp \gets \ProjKG(\hk,x)$, $H_\hk \gets \Hash(\hk, x)$, and $v\getsr \cV$;
\end{itemize}
The algorithms and properties described above are the basic algorithms for \SPHFwG. For a later use, we need to define several additional properties.

\subsection{Word Indistinguishability and Trapdoor}
First, we assume languages $\cL$ and $\cL'$ in $\cX$  are defined according to a random tape $(\sigma,\rho)$ sampled in a set $\cS_0\times\cR_0$ (i.e. from $\param \gets \setup(\sigma,\rho)$). The samplable set $\cS_0$ is defined together with its twin set $\cS_1$ such that when $\sigma \in \cS_1$, and $\param \gets \setup(\sigma,\rho)$, there exists a trapdoor $\td_\sigma$ that allows to test if a given word $x \in \cX$ is in $\cL'$ or not. We then also need the following algorithms:
\begin{itemize}
  \item $\wordgen L(\param)$: Samples and returns $x \getsr \cL$, together with its witness $w$;
  \item $\wordgen X(\param)$: Samples and returns $x \getsr \cX$; 
  \item $\wordtest(\td_\sigma, x)$, using the trapdoor $\td_\sigma$, tests if $x \in \cL'$.
\end{itemize}
As we assumed $\cL$ to be a hard subset of $\cX$ when $\sigma\in\cS_0$, we have the \textbf{Word-Indistinguishability Property}: An adversary can not distinguish between random words in $\cL$ and random words in $\cX$, for any $\sigma\in\cS_0$, with more than a negligible advantage.

The string $\sigma$ can be seen as a CRS, that admits a trapdoor when sampled from $\cS_1$. The normal use is with $\sigma\getsr\cS_0$, which needs to be efficiently samplable. When $\sigma\in\cS_1$, the trapdoor $\td_\sigma$ must be easy to compute from $\sigma$.

\subsection{Decomposition Intractability and Trapdoor}
We also define the alternate sets $\cR_1$ and $\cR'_1$ for $\cR_0$. During normal use, $\rho$ is sampled from $\cR_0$, which needs to be efficiently samplable. When $\rho \in \cR_1$, and $\param \gets \setup(\sigma,\rho)$, there exists a trapdoor $\td_\rho = (x,x',w,w')$, that must be easy to compute from $\rho$. When $\rho \in \cR'_1$, and $\param \gets \setup(\sigma,\rho)$, there exists a trapdoor $\td_\rho = (x,x')$, that must be easy to compute from $\rho$. Let us define the complement algorithm, for any $\rho\in\cR_0\cup\cR_1\cup\cR'_1$: 
\begin{itemize}
  \item $\complementword(\param,\rho,x)$: from any word $x \in \cX$, it outputs $x'$;
\end{itemize}
From this complement algorithm, we expect the following statistical properties, for any $\sigma\in\cS_0\cup\cS_1$ but $\rho\in\cR_0$:
\begin{itemize}
  \item \textbf{Complement:} for any $x\in\cX$, if $x'\gets\complementword(\param,\rho,x)$, then $x = \complementword(\param,\rho,x')$; 
  \item \textbf{Alternate:} for any $x\in \cL'$, $\complementword(\param,\rho,x)\not\in \cL'$.
\end{itemize}
But we also need a computational assumption: the \textbf{Decomposition Intractability}, which states that no adversary can generate, with non-negligible probability, for random $(\sigma,\rho)\getsr\cS_1\times\cR_0$, two words $x,y\not\in\cL'$ such that $y = \complementword(\param,\rho,x)$, and so even with the trapdoor $\td_\sigma$.

On the other hand, when $\rho\in\cR_1$, the trapdoor $\td_\rho = (x,x',w,w')$ satisfies $x$ and $x'$ are uniformly random in $\cL$ with witnesses $w, w'$, and $x' = \complementword(\param,\rho,x)$.
And when $\rho\in\cR'_1$, the trapdoor $\td_\rho = (x,x')$ satisfies $x$ and $x'$ are uniformly random in $\cL'$, and $x' = \complementword(\param,\rho,x)$.

Again, the string $\rho$ can be seen as a CRS, that admits a trapdoor when sampled from $\cR_1$ or $\cR'_1$. The normal use is with $\rho\getsr\cR_0$, which needs to be efficiently samplable. When $\rho\in\cR_1$ or $\rho\in\cR'_1$, the trapdoor $\td_\rho$ must be easy to compute from $\rho$.

Eventually, for the security proof to go through, we will make use of the \textbf{CRS Indistinguishability}: An adversary can not distinguish between $\cR_0$, $\cR_1$ and $\cR'_1$, and between $\cS_0$ and $\cS_1$, with more than a negligible advantage.

Note that we independently consider the choices between $\cS_0$ and $\cS_1$ and between $\cR_0$, $\cR_1$ and $\cR'_1$, but the latter choice could depend on the former choice. So the global CRS is the pair $\crs=(\sigma,\rho)$.

\section{Oblivious Transfer from \SPHFwG}
\label{sec:OT}
In this section we first present our construction of Oblivious Transfers based on Smooth Projective Hash Functions with Grey Zone, and then provide a security proof of our Oblivious Transfer in the Universal Composability framework

\subsection{Construction of Oblivious Transfer}
Our Oblivious Transfer uses a $\crs = (\sigma,\rho) \in \cS_0\times\cR_0$ as defined above, where we assume $\cS_0\times\cR_0 \approx \cS_1\times\cR_0 \approx \cS_0\times\cR_1 \approx \cS_0\times\cR'_1$.
We describe in Figure~\ref{fig:OTwSPHF} the \OT protocol $\varPhi^{\SPHFwG}_{\OT}$. 

\begin{figure}[!ht]
  \newcolumntype{C}{>{\centering\arraybackslash}p{1.8cm}}
  \centering
  \noindent  \fbox{\parbox{.98\linewidth}{
    \begin{tabular}{lCl}
      \multicolumn{1}{c}{Receiver} & & \multicolumn{1}{c}{Sender} \\
      with input $b\in\bit$ & & with input $m_0,m_1\in\cM$ \\
      for $\sid$ and $\crs = (\sigma,\rho)$ from $\FCRS$ & & for $\sid$ and $\crs = (\sigma,\rho)$ from $\FCRS$ \\ \hline 
      $\param \gets \setup(\sigma,\rho)$ && $\param \gets \setup(\sigma,\rho)$ \\
      $(x_b, w) \gets \wordgen L(\param)$ \\
      $x_{1-b} \gets \complementword(\rho,x_b)$ & \sendright{$x_0$}{1.5} & $x_1 \gets \complementword(\rho,x_0)$ \\
      & & for $i$ in $\{0,1\}:$ \\
      & & $\hk_i \gets \HashKG(\param)$ \\
      & & $\hp_i \gets \ProjKG(\hk_i,x_i)$ \\
      & & $H_i \gets \Hash(\hk_i, x_i)$ \\
      $H' \gets \ProjHash(c_{b,1}, x_b, w)$
        & \sendleft{$(c_0,c_1)$}{1.5}
        & $c_i=(H_i \oplus m_i, \hp_i)$ \\
      $m = H' \oplus c_{b,0}$
      \end{tabular} \\[-1em]
      \caption{General description of the protocol $\varPhi^{\SPHFwG}_{\OT}$}\label{fig:OTwSPHF}}}
\end{figure}

The protocol $\varPhi^{\SPHFwG}_{\OT}$ provides \textbf{Correctness}. Indeed, with the honest generation $(x,w) \allowbreak  \gets \wordgen L(\param)$ we have $c=(H \oplus m, \hp)$. Then, $m=H \oplus m \oplus \ProjHash(c_1, x, w)$ if and only if $H = \ProjHash(c_1, x, w)$ which is ensured due to the correctness property of the \SPHFwG.
We also need to prove the privacy. But let us proceed in the Universal Composability framework.

\subsection{Security Analysis}
The proof requires two functionalities: as our Oblivious Transfer protocol will be proven in the CRS-hybrid model (as in~\cite{C:PeiVaiWat08}), with the functionality $\FCRS$, where the two players get the same random $\crs$ from the $\sid$. In practice, as we assumed $\cS_0$ and $\cR_0$ efficiently samplable, $(\sigma,\rho)$ can be derived from $\cH(\sid)$. Then, in Figure~\ref{Fig:UC-OT}, we recall the ideal functionality for a secure oblivious transfer, where there are two first messages from the sender with $(m_0, m_1)$ and from the receiver with $b$, to initialize the process, and the final request message by the sender that decides when the receiver can get $m_b$.


\begin{figure}[t]
  \centering
  \fbox{\parbox{0.98\textwidth}{
      $\FOT$ interacts with a sender S and a receiver R:
      \begin{itemize}
      \item Upon receiving a message (\sid,\texttt{sender}, $m_0, m_1$) from S, store $(\sid, m_0, m_1)$;
      \item Upon receiving a message (\sid,\texttt{receiver}, $b$) from R, store $(\sid, b)$; 
      \item Upon receiving a message (\sid,\texttt{answer}) from the adversary, check if both records $(\sid, m_0, m_1)$ and $(\sid, b)$ exist for $\sid$. If yes, send (\sid,$m_b$) to R, and \sid{} to the adversary and halt. If not, send nothing but continue running.
      \end{itemize}
      \vspace*{-1em}
      \caption{Functionality $\FOT$\label{Fig:UC-OT}}
  }}
\end{figure} 

\begin{theorem}
  The protocol $\varPhi^{\SPHFwG}_{\OT}$ \UC-realizes $\FOT$ in the $\FCRS$-hybrid model in the static-corruption setting, from any \SPHFwG.
\end{theorem}
We stress that we consider static corruptions only, where the corrupted players are known when each protocol execution starts.
\begin{proofgame}
  \item This is the real game, where $\FCRS$ samples $\crs$ in $\cS_0 \times \cR_0$.
  
  \item In this game, the simulator $\cS$ simulates itself the sampling of $\crs = (\sigma,\rho) \getsr \cS_0\times\cR_0$, and generates correctly every flow from the honest players, as they would do themselves, knowing the inputs $(m_0, m_1)$ and $b$ sent by the environment to the sender and the receiver, respectively.
  
  \item In this game, we deal with \textbf{corrupted receivers}. Instead of sampling $\crs = (\sigma,\rho) \getsr \cS_0\times\cR_0$, the simulator $\cS$ samples $\crs = (\sigma,\rho) \getsr \cS_1\times\cR_0$, and therefore with the trapdoor $\td_\sigma$. This game is indistinguishable from the previous one due to the \emph{CRS Indistinguishability}.
  
  \item In this game, the simulator $\cS$ uses the trapdoor $\td_\sigma$ to get $t_i = \wordtest(x_i, \td_\sigma)$ for $i\in\{0,1\}$. If $t_0=t_1=0$ (none of the words are in $\cL'$), $\cS$ aborts. 
  This game is indistinguishable from the previous one, under the \emph{Decomposition Intractability}, as $(\sigma,\rho)\in\cS_1\times\cR_0$.
  
  \item If $t_0=t_1=0$, we still abort. If $t_0=t_1=1$ we set $b=0$, otherwise, we set $b$ such that $t_b=0$. Next, the simulator $\cS$ proceeds on $m_b$ with $x_b$ and on a random message with $x_{1-b}$. Under the smoothness of the $\SPHFwG$, as $x_{1-b}\in\cL'$, and the \emph{One-Time Pad Semantic Security}, this game is statistically indistinguishable from the previous one. 
  
  \item In this game, we deal with \textbf{corrupted senders}. Instead of sampling $\crs = (\sigma,\rho) \getsr \cS_0\times\cR_0$, the simulator $\cS$ samples $\crs = (\sigma,\rho) \getsr \cS_0\times\cR_1$, and therefore with the trapdoor $\td_\rho = (x,x',w,w')$. This game is indistinguishable from the previous one due to the \emph{CRS indistinguishability}.
  
  \item In this game, the simulator $\cS$ respectively sets $(x_0,w_0,x_1,w_1)$ as $(x,w,x',w')$ from $\td_\rho$. It can then retrieve both $m_0$ and $m_1$. This game is indistinguishable from the previous one due to the \emph{Word Indistinguishability}, and the uniform distribution of the trapdoor.
  
  \item We now deal with \textbf{honest players}.
  Instead of sampling $\crs = (\sigma,\rho) \getsr \cS_0\times\cR_0$, the simulator $\cS$ samples $\crs = (\sigma,\rho) \getsr \cS_0\times\cR'_1$, and therefore with the trapdoor $\td_\rho = (x,x')$, and simulates the flows with random $m_0,m_1\getsr\cM$ and random $b\getsr\bit$. Under the \emph{CRS Indistinguishability} and the smoothness of the \SPHFwG, as both $x,x'\in\cL'$, coupled with the \emph{One-Time Pad Semantic Security}, this game is indistinguishable from the previous one.
  
  \item This is the ideal game 
  We can now make use of the functionality $\FOT$ which leads to the following simulator:
  \begin{itemize}
    \item If no participant is corrupted, one uses $\crs\getsr\cS_0\times\cR'_1$, and the simulator $\cS$ simply uses random inputs for the sender and the receiver;
    \item If the receiver is corrupted, one uses $\crs\getsr\cS_1\times\cR_0$, and the simulator $\cS$ extracts $b$ using the trapdoor $\td_\sigma$, and sends $(\sid, \textsf{receiver}, b)$ to $\FOT$;
    \item If the sender is corrupted, one uses $\crs\getsr\cS_0\times\cR_1$, and the simulator $\cS$ extracts $m_0,m_1$ using the trapdoor $\td_\rho$, and sends $(\sid, \textsf{sender}, m_0, m_1)$ to $\FOT$;
    \item The adversary sends $(\sid, \textsf{answer})$ when it decides to deliver the result to the receiver.
  \end{itemize}
\end{proofgame}

\subsection{Noisy Homomorphic Encryption Setup} \label{subsec:OT:Setup}
We now define a general setup leading to an instanciation of our Oblivious Transfer from any (possibly Noisy) Homomorphic Encryption.

We consider a noisy encryption scheme $\varPi=(\setup, \keygen, \allowbreak \encrypt, \allowbreak \decrypt)$ with possible decryption failures.
$\cX$ as the ciphertext space of $\varPi$, whereas
\begin{align*}
  \cL = \{\encrypt(\pk, 0; r) \} \subset \cX \text{ and } \cL' = \{c \in \cX, \decrypt(\sk, c) \neq 0 \} \subset \cX.
\end{align*}
Sets $\cS_0$ and $\cS_1$ can both be seen as public keys $\pk$ generated from $\keygen(1^\kappa)$ except that when $\sigma \in \cS_1$, the secret key $\sk$ is known and defines the trapdoor $\td_\sigma$.
Hence, $\sigma$ (which defines the public key $\pk$) defines the sets $\cL$ and $\cL'$ in $\cX$.
On the other hand, we can define $\cR_0=\cX$, the set of all the ciphertexts, or a superset, with uniform distribution; $\cR_1=\{ c_0 \otimes c_1 \}$, for two ciphertexts $c_0, c_1$ in $\cL$, following the distribution of the encryption algorithm, on plaintext 0, and according the distribution of the randomness $r_0,r_1$, which allows to define the trapdoor $\td_\rho$ as $(c_0,c_1,r_0,r_1)$;
$\cR'_1=\{ c_0 \otimes c_1 \}$, for two ciphertexts $c_0, c_1$ in $\cL'$, following the distribution of the encryption algorithm, on non-zero plaintexts, which allows to define the trapdoor $\td_\rho$ as $(c_0,c_1)$.
The setup defined above verifies both basic assumptions required to make the Oblivious Transfer Universally Composable:
\begin{itemize}
  \item \emph{CRS Indistinguishability}: Under the \emph{semantic security} of the encryption scheme $\varPi$, $\cL$, $\cL'$, and $\cX$ are indistinguishable. The homomorphic property implies that $\{x \otimes x' | (x,x') \in \cX^2 \}=\cX$. As a consequence, we have indistinguishability between $\cR_0 = \cX = \{x \otimes x' | (x,x') \in \cX^2 \}$, $\cR_1 = \{x \otimes x' | (x,x') \in \cL^2 \}$, and $\cR'_1 = \{x \otimes x' | (x,x') \in {\cL'}^2 \}$. Furthermore, as $\cS_0=\cS_1$, they are perfectly indistinguishable;
  \item \emph{Word Indistinguishability}: Under the \emph{semantic security} of the encryption scheme $\varPi$, one can not distinguish between $c_0 \in \cL$, an encryption of 0 and $c_1 \in \cX$, an encryption of a random value.  
\end{itemize}
Additional properties will depend on concrete instantiations.

\section{Concrete Instantiations of $\SPHFwG$}
\label{sec:instantiations}
We now provide two concrete instantiations of $\SPHFwG$ based on the Diffie-Hellman and Learning With Errors problems.
As both constructions rely on an Homomorphic Encryption scheme, we can already consider the basic properties shown in Section \ref{subsec:OT:Setup}. 
\subsection{Instantiation from the Diffie-Hellman Problem} \label{sec:instantiation:dh}
In this section, we focus on elliptic curve based cryptography, using the Decisional Diffie-Hellman assumption in a prime-order group. 

\begin{definition}[Decisional Diffie-Hellman (DDH)]
	In a group $\bG$ of prime order $p$, the \textit{Decisional Diffie-Hellman} problem consists in, given $g^a$ and $g^b$, distinguishing $g^{ab}$ from $g^c$, for $a,b,c\getsr\bZ_p$.
\end{definition}


\subsubsection{ElGamal Encryption.}
\label{subsec:CPA_enc:EG}
As expected above, we need an IND-CPA (a.k.a. with semantic security) encryption scheme, with homomorphism.
We use the ElGamal encryption scheme \cite{ElGamal85} in a group $\bG = \langle g \rangle$ of prime order $p$, defined by the $\setup$ algorithm:
\begin{itemize}
	\item $\keygen(1^\kappa)$: picks $\beta \getsr \bZ_p$, and sets $\pk=h=g^\beta$, $\sk=\beta$.
	\item $\encrypt(\pk = h = g^\beta,\; M \in \bG)$ encrypts the message $M$ under the public key $\pk$ as follows: Pick $r \getsr \bZ_p$;
	Output the ciphertext: $c = (g^r, h^r \cdot M)$;	
	\item $\decrypt(\sk, c = (c_0,c_1))$ decrypts the ciphertext $c$ using the decryption key $\sk$ as follows: $M= c_1 / c_0^\sk$.
\end{itemize}

\begin{theorem}
	\label{thm:CPA_security:EG}
	The above ElGamal encryption scheme is IND-CPA under the Decisional Diffie-Hellman assumption.
\end{theorem}

\subsubsection{\SPHFwG from ElGamal Encryption.}
From the above ElGamal encryption scheme $\EG=(\setup,$ $\keygen,\encrypt, \decrypt)$, in a group $\bG$, denoted multiplicatively, of prime order $p$, with generator $g$.

We set $\cS_0=\cS_1=\{ \sigma = h = g^{\td_\sigma}; {\td_\sigma} \getsr \bZ_p\}$.
Then, $\cR_0$ is defined as $\bG^2 = \{\rho = (\hg, \hh) \getsr \bG^2\}$
and $\cR_1$ as $\{\rho = (\hg = g^{\td_\sigma}, \hh = h^{\td_\sigma}); \hg\getsr \bG, \td_\sigma \getsr \bZ_p \}$. 
The $\crs$ is set as $(\sigma,\rho)$. One can note that witnesses only exist when $\rho\in\cR_1$, then $\td_\rho = ((c_0,c_1), (c_0\cdot\hg,c_1\cdot\hh),r, r+\td_\sigma)$, where $(c_0,c_1)$ is an encryption of $M = g^0$, with randomness $r$; while $\td_\sigma$ always exists, it is not necessarily known.

From the above generic construction, we have $\cX= \{ (g^r, h^r \cdot  M) , M \in \bG\} = \bG^2$ and $\cL=\{(g^r , h^r)\}$, which are indistinguishable under the Decisional Diffie-Hellman assumption. With $\param = (g, \sigma = h)$, which determines all the sets (specified by the $\setup$ algorithm), we can define:
\begin{itemize}
	\item $\hk = \HashKG(\param) = (\alpha,\beta) \getsr \bZ_p^2$;
	\item $\hp = \ProjKG(\hk)=g^\alpha h^\beta$;
	\item $H = \Hash(\hk, x=(u,v))=u^\alpha v^\beta \in \bG$;
	\item $H' = \ProjHash(\hp,x,w=r)=\hp^r$, if $x = (g^r, h^r)\in \cL$.
\end{itemize}
This is a word-independent \SPHFwG. And we can show the expected properties:
\begin{itemize}
	\item \textbf{Correctness:} When $x = (u,v) = (g^r,h^r) \in\cL$, with witness $r$, $H = u^\alpha v^\beta = (g^\alpha h^\beta)^r = \hp^r = H'$;
	\item \textbf{Smoothness:} When $x = (u,v) = (g^r,h^{r'}) \not\in\cL$, then $r' = r + r''$ with $r''\neq 0$: $H = u^\alpha v^\beta = (g^\alpha h^\beta)^r \times g^{r'' \beta} = \hp^r \times g^{r'' \beta} = H' \times g^{r'' \beta}$. But $\beta$ is perfectly hidden in $\hp$, and $g^{r'' \beta}$ is perfectly unpredictable;
	\item \textbf{Decomposition Intractability:} We can note that in ElGamal encryption there is no decryption failure: all the ciphertexts can be covered by the encryption algorithm, and the decryption perfectly inverts the encryption process. So $\cL' = \cX \backslash \cL$. A random ciphertext $\rho$ encrypts an $M \neq 1$ with overwhelming probability. Then, when it encrypts $M\neq 1$, from the homomorphic property, this is impossible to have two encryptions of 1 whose product is $\rho$. Hence, the \emph{decomposition intractability} is statistical: the probability of existence of the decomposition is bounded by $1/p$, on $\rho$, even knowing the decryption key, and thus the trapdoor $\td_\sigma$.
  \end{itemize}
  Note that this construction exactly corresponds to the one from~\cite{C:CraSho98}.

\subsection{Instantiation from the Learning With Errors Problem} \label{sec:instantiation:lwe}
In this section, we focus on lattice-based cryptography. We are going to show how to instantiate the various required components from LWE:

\begin{definition}[Shortest Independent Vectors Problem (SIVP$_\gamma$)]

The approximation version SIVP$_\gamma$ is the approximation version of SIVP with factor $\lambda$. Given a basis $\textbf{B}$ of an $n$-dimensional lattice, find a set of $n$ linearly independent vectors $v_1, \ldots , v_n \in \mathcal{L}(\textbf{B})$ such that $\|v_i\| \leq \gamma(n)\cdot \lambda_n(\textbf{B})$.
for all $1 \leq i \leq n$. The approximation factor $\gamma$ is typically a polynomial in $n$, the non approximated version assumes $\gamma = 1$.
\end{definition}

\begin{definition}[Learning With Errors (LWE)]
	Let $q \geq 2$, and $\chi$ be a distribution over $\bZ$.
	The \textit{Learning With Errors} problem LWE$_{\chi, q}$ consists in, given a polynomial number of samples, distinguishing the two following distributions:
	\begin{itemize}
		\item $(\mathbf a, \langle \mathbf a, \mathbf s \rangle + e)$, where $\mathbf a$ is uniform in $\bZ_q^n$,  $e \gets \chi$, and $\mathbf s \in \bZ_q^n$ is a fixed secret chosen uniformly, and where $\langle \mathbf a, \mathbf s \rangle$ denotes the standard inner product.
		\item $(\mathbf a, b)$, where $\mathbf a$ is uniform in $\bZ_q^n$, and $b$ is uniform in $\bZ_q$.
	\end{itemize}
\end{definition}

\subsubsection{Regev Encryption.}
Regev~\cite{STOC:Regev05} showed that for $\chi = D_{\bZ, \sigma}$, a Gaussian centered distribution in $\bZ$ for any standard deviation $\sigma \geq 2 \sqrt{n}$, and $q$ such that $q/\sigma = \textsf{poly}(n)$, LWE$_{\chi, q}$ is at least as hard as solving worst-case SIVP for polynomial approximation factors, which is assumed to be hard to solve, even for quantum computers.

\subsubsection{Trapdoor for LWE.} \label{sec:lwe-enc-trap}

Throughout this paper, we will use the trapdoors introduced in \cite{EC:MicPei12} to build our public matrix $\mathbf A$. Define $g_{\mathbf A}(\mathbf s, \mathbf e) = \mathbf A \mathbf s + \mathbf e$, the gadget matrix $\mathbf G$ as $\mathbf G^t = \mathbf I_n \otimes \mathbf g^t$, where $\mathbf g^t = [1,2,\dots, 2^{k} ]$ and $k = \lceil \log q \rceil -1$, and let $\mathbf H \in \bZ_q^{n\times n}$ be invertible.
The notation $[ \mathbf A \,|\, \mathbf B]$ is for horizontal concatenation, while $[ \mathbf A \,;\, \mathbf B]$ is for vertical concatenation.

\begin{lemma}[{\cite[Theorems 5.1 and 5.4]{EC:MicPei12}}] \label{trapdoor}
	There exist two PPT algorithms $\TrapGen$ and $g^{-1}_{(\cdot)}$ with the following properties assuming $q \geq 2$ and $m \geq \Theta(n	 \log q)$:
	\begin{itemize}
		\item $\TrapGen(1^n,1^m,q)$ outputs $(\mathbf T, \mathbf A_0)$, where the distribution of the matrix $\mathbf A_0$ is at negligible statistical distance from uniform in $\bZ_q^{m\times n}$, and such that $\mathbf T \mathbf A_0 = \mathbf 0$, where $s_1(\mathbf T) \leq O(\sqrt m)$ and where $s_1(\mathbf T)$ is the operator norm of $\mathbf T$, which is defined as $\max_{\mathbf x \neq 0} \|\mathbf T \mathbf x\| / \|\mathbf x\|$.\footnote{The bound on $\mathbf s_1(\mathbf T)$ holds except with probability at most $2^{-n}$ in the original construction, but we assume the algorithm restarts if it does not hold.}
		\item Let $(\mathbf T, \mathbf A_0) \gets \TrapGen(1^n,1^m,q)$. Let $\mathbf A_{\mathbf H} = \mathbf A_0  + [\mathbf 0 \,; \, \mathbf G \mathbf H]$ for some invertible matrix $\mathbf H$  called a \emph{tag}. Then, we have $\mathbf T \mathbf A_{\mathbf H} = \mathbf G \mathbf H$. Furthermore, if $\mathbf x \in \bZ_q^m$ can be written as $\mathbf A_{\mathbf H} \mathbf s + \mathbf e$, with $\mathbf s \in \bZ_q^n$ and $\mathbf e \in \bZ_q^m$ where $\|\mathbf e\| \leq B' \defeq q / \Theta(\sqrt m)$, then $g_{\mathbf{ A_{\mathbf H}}}^{-1}(\mathbf T, \mathbf x, \mathbf{H})$ outputs $(\mathbf s,\mathbf e)$.
	\end{itemize}
	
\end{lemma}
More precisely, to sample $(\mathbf T, \mathbf A_0)$ with $\TrapGen$, we sample a uniform $\mathbf {\bar A} \in \bZ_q^{\bar m \times n}$ where $\bar m = m - nk = \Theta(n\log q)$, and some $\mathbf R \gets \mathcal{D}^{nk \times \bar m}$, where the distribution $\mathcal{D}^{nk \times \bar m}$ assigns probability $1/2$ to $0$, and $1/4$ to $\pm 1$. We output $\mathbf T = [ -\mathbf R \,| \, \mathbf I_{nk} ]$ along with $\mathbf A_0 = [ \mathbf {\bar A} \,;\, \mathbf R \mathbf {\bar A}]$. Then, given a tag $\mathbf H$, with $\mathbf A_{\mathbf H} = \mathbf A_0  + [\mathbf 0 \,; \, \mathbf G \mathbf H]$, we have: $\mathbf T \mathbf A_{\mathbf H} = \mathbf G \mathbf H$.

We will only consider a fixed tag $\mathbf H= \mathbf I$, for the Micciancio-Peikert encryption \cite{EC:MicPei12}. Our construction only requires CPA encryption so we don't need several tags, but we need to be able to reject improperly computed ciphertexts, and the gadget matrix is here, to allow this extra control during the decryption.

\subsubsection{LWE Encryption \`a la Micciancio-Peikert.}
\label{subsec:CPA_enc:lwe}

For this scheme, we assume $q$ to be an odd prime. We set an encoding function for messages $\Encode(\mu \in \{0,1\}) = \mu \cdot (0, \dots 0, \lceil q/2 \rceil)^t$. Note that $2 \cdot \Encode(\mu) = (0, \dots, 0, \mu)^t \bmod q$, as $\lceil q/2 \rceil$ is the inverse of $2 \bmod q$, for such an odd $q$.

Let $(\mathbf T, \mathbf A_0) \gets \TrapGen(1^n,1^m,q)$. The public encryption key is $\pk = \mathbf A_0$, and the secret decryption key is $\sk = \mathbf T$. 

\begin{itemize}
	\item $\encrypt(\pk = \mathbf A_0,\; \mu \in \{0,1\})$ encrypts the message $\mu$ under the public key $\pk$ as follows: Let $\mathbf A = \mathbf A_0 + [\mathbf 0 \,;\, \mathbf G]$. Pick $\mathbf s \in \bZ_q^n$, $\mathbf e \gets D^{m}_{\bZ, t}$ where $t =  \sigma \sqrt{m} \cdot \omega(\sqrt{\log n})$. Restart if $\|\mathbf e\| > B$, where $B \defeq 2t \sqrt m$.\footnote{This happens only with exponentially small probability $2^{-\Theta(n)}$.}  Output the ciphertext:
	\[ \mathbf c = \mathbf A \mathbf s + \mathbf e + \Encode(\mu) \bmod q \enspace. \]
	
	\item $\decrypt(\sk = \mathbf T,\; \mathbf c \in \bZ_q^m)$ decrypts the ciphertext $\mathbf c$ using the decryption key $\sk$ as follows: With $B'' \defeq q / 2\Theta(\sqrt m)$, output \\
	\[\begin{cases}
		\mu &\text{if } g^{-1}_{\mathbf A}(\mathbf T, 2\mathbf c, \mathbf I) = (2\mathbf s, 2\mathbf e + (0, \dots ,0 , \mu)) \\
		& \qquad \text{ where } \mathbf s \in \bZ^n_q, \mathbf e \in \bZ^m 
		\text{ and } \|\mathbf e\| \leq B'' \enspace, \\ 
		\bot &\text{otherwise.}\footnotemark
	\end{cases}\]
\end{itemize}
Noting $\Lambda(A)= \{\textbf{As} | \textbf{s} \in \bZ_q^n\}$, honestly generated ciphertext $\textbf{c}$ are such that $d(\mathbf c-\Encode(\mu),\Lambda(\textbf{A}))\leq B$, while the decryption procedure is guaranteed not to return $\mu$ as soon as $d(\mathbf c-\Encode(\mu),\Lambda(\textbf{A}))> B''$.
\footnotetext{Note that the inversion algorithm $g^{-1}_{(\cdot)}$ can succeed even if $\|\mathbf e\| > B''/2$, depending on the randomness of the trapdoor. It is crucial to reject decryption nevertheless when $\|\mathbf e\| > B''$ to ensure security.}
From the decryption procedure, we have:
\begin{equation*}
	\mu' \defeq \decrypt(\mathbf T, \mathbf c) \neq \bot  \quad \Longleftrightarrow \quad d(\mathbf c - \Encode(\mu'), \Lambda(\mathbf A)) < B''\enspace.
\end{equation*}
Suppose that $m \geq \Theta(n\log q)$. The scheme is correct as long as $B \leq B''$, or equivalently 
$2 \sigma m^{3/2} \cdot \omega(\sqrt{\log n}) \leq q$.

\begin{theorem}
	\label{thm:CPA_security:LWE}
	Assume $m \geq \Theta(n \log q)$. The above scheme is IND-CPA assuming the hardness of the LWE$_{\chi, q}$ problem for $\chi = D_{\bZ, \sigma}$.
\end{theorem}
Furthermore, this encryption scheme is homomorphic for plaintexts in $(\bZ_2,+)$, and ciphertexts in $\bZ_q^m$ with component-wise addition.

\subsubsection{Bit-\SPHFwG from LWE Encryption Scheme.}
We consider, an LWE encryption scheme defined with a superpolynomial modulus.
More precisely, we set $m = n \log(q), t = \sqrt{mn} . \omega(\sqrt{\log(n)})$, $k = \Theta(n), s \geq \Theta(\sqrt{n}) \wedge s/q = \textsf{negl}(n), s= \Omega(m k^2q^{2/3})$.
We also set $R$ to be a \emph{probabilistic} rounding function from $[0,1]$ to $\{0,1\}$, such that $R(x) = 1$ with probability $0.5 \cdot \cos(\frac{2 \pi x}{q})$ and 0 otherwise.

We set $\cS_0=\cS_1=\{ \sigma = \mathbf A = \mathbf A_0 + [\mathbf 0 \,;\, \mathbf G] | (\mathbf T, \mathbf A_0) \gets \TrapGen(1^n,1^m,q)\}$, $\td_\sigma$ being $\mathbf T$.
Then, $\cR_0$ is defined as $\{\rho = \mathbf v \in \bZ_q^m  \}$
and $\cR_1$ is the set composed of all the sums of two honest encryptions of 0, in other words
$\{\rho= \mathbf A (\mathbf s+\mathbf{s'}) + \mathbf e + \mathbf e' \bmod q \enspace | \mathbf s, \mathbf s' \in \bZ_q^n$, $\mathbf {e,e'} \gets D^{m}_{\bZ, t} \wedge \|\mathbf e\| \leq B\wedge \|\mathbf e'\| \leq B  \}$ with $\td_\rho=(\mathbf A \mathbf s + \mathbf e, \mathbf A \mathbf s' + \mathbf e', (\mathbf s,\mathbf e), (\mathbf s', \mathbf e'))$.

With $\cX= \{ \mathbf c \getsr \bZ_q^m \}$, $\cL=\{\mathbf c | \exists \mathbf  s, \mathbf e, \mathbf c= \encrypt(\textbf{A}_0,0;\mathbf s,\mathbf e)\}$ defined following the description above, and $\cL' = \{ \mathbf c \in\cX | \decrypt(\textbf{T},\mathbf c) \neq 0\}$,  note that $\mathbf s$ could be enough as a witness for $\mathbf c = \mathbf A \mathbf s + \mathbf e \in\cL$, as one can check $\mathbf e = \mathbf c - \mathbf A \mathbf s$ is small enough. This defines the $\setup$ algorithm, and we have:
\begin{definition}[Bit-\SPHFwG over Micciancio-Peikert like Ciphertexts \cite{PKC:BBDQ18}]
	For $k=\Theta(n)$, and picking $s\geq {\Theta}(\sqrt{n})$, and $s=\Omega(mk^2q^{2/3})$, we can define:
\begin{itemize}
	\item $\HashKG(\param)=\hk=\mathbf{h} \getsr D^m_{\mathbb{Z},s}$
	\item $\ProjKG(\hk)=\hp = \mathbf{A}^t \mathbf{h}$
	\item $\Hash(\hk, \mathbf c)=R(\langle \hk,\mathbf c \rangle) = R(\langle \mathbf{h},\mathbf c \rangle) \in \{0,1\}$
	\item $\ProjHash(\hp,\mathbf c,w=\mathbf s)=R(\langle \hp,\mathbf s \rangle)=R(\langle \mathbf{A}^t \mathbf{h},\mathbf s \rangle)$
\end{itemize}
\end{definition}
For a word $\mathbf c = \mathbf A \mathbf s+ \mathbf e$ in the language $\cL$, $\langle \mathbf{h},\mathbf c \rangle$=$\mathbf{h}^t \mathbf{A} \mathbf{s} + \mathbf{h}^t \mathbf{e} = \langle \mathbf{A}^t \mathbf{h},\mathbf s \rangle +  \mathbf{h}^t \mathbf{e}$.
And by construction $\mathbf{h}^t \mathbf{e}$ is small. The choice of the rounding function $R(x)$, characterized by a coin flip where the outcome 1 is weighted by $0.5 \cdot \cos(\frac{2 \pi x}{q})$, is such that it allows to cancel out this small noise most of the time, while providing smoothness for words outside the language (ensuring that $R(\langle \hk,\mathbf c \rangle)$ is random when given only $\hp$) 


It was shown in \cite{PKC:BBDQ18}, that for this choice of random function, such bit-\SPHFwG achieves negligible-universality, thanks to the rounding function, but $(3/4 + o(1))$-correctness for the chosen set of parameters.

\subsubsection{Full-Fledged \SPHFwG from LWE.}
The previous construction has limitations as it is neither perfectly correct, nor smooth, we need to apply a transformation to reach those goals. This transformation is explained below, first informally, then in more details:
\begin{itemize}
	\item It is a bit-function meaning the final hash value lives in $\{0,1\}$, while one needs a larger mask. To solve this issue, one has to run it in parallel a linear number of times, to have an output string long enough.
	\item The correctness is imperfect. The output bit only matches with probability $3/4 + o(1)$. As such, applications running $\encrypt(\pk, m; r)$ should encryption a redundant version of $m$, with an error-correcting code, $\mathsf{ECC}(m)$. Such transformation makes the \SPHF word-dependent (i.e. the projection key is dependent on the user/receiver input), however in our scenario, such a word-dependent function is enough.
\end{itemize}
More formally, given a word $\mathbf c\in\cX$, for any $\ell = \Omega(n)$ an error-correcting code $\mathsf{ECC}$ capable of correcting $\ell/4$ errors, then, we can define the \SPHF as:
%
\begin{itemize}
	\item $\mathsf{SETUP}(1^\kappa)$: Outputs the result from $\setup(1^\kappa)$
	\item $\mathsf{HASHKG}(\param)$: Picks a random values $K \gets \{0,1\}^\kappa,$ and $\forall i \in [\ell],$ gets $\hk_i= \HashKG(\param)$,  and set $\mathsf{HK}=(\{\hk_i\},K)$;
	\item $\mathsf{ProjKG}(\mathsf{HK},\mathbf c): \forall i \in [\ell],$ gets $\hp_i=\ProjKG(\hk_i), H_i = \Hash(\hk_i,\textbf{c})$.  It then computes $T=\mathsf{ECC}(K) \oplus S$ where $S = (H_i)_{i\in [\ell]}$, and outputs $\mathsf{HP}=((\hp_i)_{i\in [\ell]}, T)$;
	\item $\mathsf{HASH}(\mathsf{HK},\textbf{c})$: Returns $K$, from $\mathsf{HK}$;
	\item $\mathsf{PROJHASH}(\mathsf{HP},\textbf{c}, w=\textbf{s}): \forall i \in [\ell],$ computes $H'_i=\ProjHash(\hp_i,\textbf{c},\textbf{s})$. Then computes $S'=(H'_i)_{i\in [\ell]}$, and finally $K' = \mathsf{ECC}^{-1}(T \oplus S')$.
\end{itemize}
Such transformation allows to achieve \textit{smoothness} which can be proven with an hybrid argument, handling intermediate distributions where the first $H_i$ values are random. 
The \textit{correctness} is simply inherited from the correcting-code capacity, while the number of errors to be corrected can be estimated thanks to the Hoeffding's bound~\cite{Hoeffding63}.
We can guarantee the expected properties:
\begin{itemize}
	\item \textbf{Correctness:} When $x = \mathbf c \in\cL$, with the above conversion, we have $K = K'$ with overwhelming probability, thanks to the error-correcting code;
	\item \textbf{Smoothness:} When $x = \mathbf c \not\in\cL$, then the value $K$ is random from an adversary point of view, as the parallelization technique allows to transform the negligible-universality to a classical smoothness (at the cost of a word-dependent \SPHF);
	\item \textbf{Half Decomposition Intractability:} A random vector $\rho$ should not be split into two ciphertexts that could be decrypted to 0, or at least not too often. We first deal with \emph{half} decomposition intractability, when at most half of the random vectors can be split.
	To get a lower-bound on the number of vectors like such $\rho$, we can remark that a vector verifies this property as soon as $d(\rho,\Lambda(\mathbf A))$ is greater than 2 times the decryption bound.
	
	This is the reason, why we took a conservative value $B'' = B'/2$ in the encryption compared to classical Micciancio-Peikert encryption. By halving the decryption radius, we ensured that adding two elements that still decrypt within this bound will fall on classically decryptable ciphertexts. As such, at least half the elements cannot be reached (those that classically decrypted to 1). 
	Hence, $\Pr_{\rho \in \bZ_q^m}[\exists \mathbf c,\mathbf d | \rho = \mathbf c  + \mathbf d \wedge \decrypt(\sk, \mathbf c)=\decrypt(\sk,\mathbf d)=0] \leq 1/2$. This is a statistical bound, that holds even when knowing the decryption key.
\end{itemize}
Another amplification is required to make \emph{full-fledged decomposition intractability}, by working on ciphertexts $(\mathbf{c}_j)_{j\in[k]}$, with $k$ parallel executions of the \SPHFwG, with a final XOR of all the outputs, so that the smoothness for one word is enough to get the smoothness for the vector of words, but the correctness on all the words leads to the global correctness. The acceptable language, for correctness is then:
\begin{align*}
	\tilde{\cL} & = \cL^k = \{(\mathbf{c}_j)_{j\in[k]} | (\forall j\in[k]), \exists (\mathbf{s}_j, \mathbf {e}_j), \mathbf{c}= \encrypt(\textbf{A}_0,0;\mathbf{s}_j,\mathbf{e_j})\}\subset\cX^k
\end{align*}
whereas the language for the smoothness becomes:
\begin{align*}
	\tilde{\cL}' & = {\cL'}^k = \{(\mathbf{c}_j)_{j\in[k]} | (\exists j\in[k]), \decrypt(\textbf{T},\mathbf{c}_j) \neq 0\}\subset\cX^k
\end{align*}
Then, for random $(\rho_j)_{j\in[k]} \getsr \cX^k$, a decomposition would be a list of pairs
$(\mathbf{c}_j,\mathbf{d}_j)_{j\in[k]} \in (\cX \times \cX)^k$ such that for all $j$,
$\rho_j = \mathbf{c}_j + \mathbf{d}_j$ and $\decrypt(\textbf{T}, \mathbf{c}_j)=\decrypt(\textbf{T},\mathbf{d}_j)=0$, which only exists with probability less than $1/2^k$. 
We thus have achieved all the security properties required for our applications.

\section{Conclusion}
\label{sec:conclusion}
In this paper, we introduced \emph{Smooth Projective Hash Functions with Grey Zone}, that generalize \SPHF to language subjected to gaps, thanks to the \emph{Decomposition Intractability} property. 
This is enough to get Oblivious Transfer proven secure in the Universally Composable model. As such a primitive can be obtained from the LWE problem, we can then obtain a UC-secure post-quantum Oblivious Transfer.

\bibliographystyle{alpha}
\bibliography{biblio,cryptobib/abbrev3,cryptobib/crypto}

\end{document}